# Bias-Modulated Electronically-Tunable Optoelectronic Chromatic Dispersion in Ge PN Photodiodes

*Uttama Kumar Saint[1]*, Ziv Glasser[1], Egor Liokumovitch[2], Endalamaw Ewnu Kassa[1], Asaf Albo[1], Shmuel Sternklar[1,2]*

[1]Department of Electrical and Electronic Engineering, Ariel University, Ariel 40700, Israel

[2]PerCiv Ltd., Gazit 1934000, Israel

*Uttama_k.saint@msmail.ariel.ac.il

## Abstract

Optoelectronic Chromatic Dispersion (OED) is a source of significant wavelength-dependent current in photodiodes, offering a pathway to computational spectroscopy in a photodiode-only platform without the need for additional optical or mechanical components. Previous studies have harnessed OED through light modulation. Here, we present a theoretical and experimental investigation of OED driven solely by bias-modulation: the bias voltage on a Ge PN photodiode is modulated sinusoidally, and the optical illumination in the C- and L-bands is held constant. The spectral response is due to wavelength-dependent optical absorption together with bias-dependent competition between carrier drift, displacement current, and mobile-charge screening. With the RF phase-shift technique, the maximum measured OED sensitivity is +4.13 deg/nm in the L-band, significantly larger than that of light-modulated OED in Ge. The 15 mdeg phase-noise floor yields a spectral resolution of 3.6 pm. These results establish bias-modulated OED as an electronically reconfigurable dispersive mechanism for applications in chip-scale computational spectrometers, photonic integrated circuits (PICs), on-chip RF photonics, and wavelength-division multiplexing.

**Keywords:** Optoelectronic Chromatic Dispersion, Bias modulation, Ge Photodiodes, optical spectroscopy

## 1. Introduction

The ubiquitous photodetector plays a critical role in nearly all areas of science and technology. They are central to a wide range of optoelectronic technologies, including optical communication, imaging, biosensing, and wearable devices [1]. As applications increasingly demand devices that are energy-efficient, low-cost, and capable of detecting across a wide spectral range, the development of photodetectors with broadband responses and high sensitivity has become an important research priority.

In addition to basic photoelectric conversion, photodetectors exhibit wavelength-dependent carrier transport. This phenomenon arises from the spectral dependence of the semiconductor absorption coefficient $\alpha(\lambda)$, which governs the depth of photon penetration into the device. Because the carrier transit time is directly linked to the absorption depth, incremental changes in the wavelength can produce measurable variations in carrier transport. As illustrated in Fig. 1, a slight increase in the wavelength of illumination causes its penetration depth to either approach the junction (for sufficiently short wavelengths, $\lambda_1$) or recede from the junction (for longer wavelengths, $\lambda_2$). When photons are absorbed within the intrinsic region, they generate electron-hole pairs. Under the applied electric field, electrons and holes drift in opposite directions toward their respective contacts, contributing to the drift current. Therefore, the wavelength-dependent absorption profile determines the spatial distribution of photocarrier generation within the intrinsic region and the distances that the generated carriers travel before collection. This results in either a reduction or an increase in the average charge carrier transit time and phase delay of the modulated photocurrent. In this manner, each wavelength component of the illumination generates electron–hole pairs at different depths, and their subsequent drift and diffusion contribute to the RF amplitude and phase $\Omega\tau$ of the photocurrent, where $\Omega = 2\pi f$ is the angular frequency, and $\tau$ is the effective carrier transit time. This effect, known as Optoelectronic Chromatic Dispersion (OED) [2,3], is the basis for photodiode-only, compact, high-resolution wavelength sensors and spectroscopic tools. By sweeping across the RF frequency range, the complete optical spectrum can be reconstructed [4–8].

A change in optical wavelength $\Delta\lambda$, changes the effective carrier transit time, $\Delta\tau$. This wavelength dependance is characterized by the OED dispersion coefficient, $\widehat{D}_{OED} = \frac{\Delta\tau}{\Delta\lambda}$, so a change in optical wavelength produces a modulation phase delay $\Delta\theta = \Omega\widehat{D}_{OED}\Delta\lambda$ . The OED sensitivity, defined as the phase change per unit wavelength change, is $S_{OED} = \frac{\Delta\theta}{\Delta\lambda} = \Omega\widehat{D}_{OED}$ [2,3]. As explained above, and with reference to Fig. 1, the sign and magnitude of the OED sensitivity depend on the location of carrier generation relative to the junction. The magnitude of OED sensitivity is proportional to $\alpha^{-1}\frac{d\alpha}{d\lambda}$, i.e., penetration depth and the slope of the absorption dispersion [3]. The magnitude of the OED dispersion coefficient is enormous over wide spectral regions, comparable to the dispersion exhibited by hundreds of kilometers of optical fibers under the same modulation frequencies [9]. The OED response can be enhanced, suppressed, or tuned for specific applications through appropriate choices of materials, intrinsic layer thickness, doping [10], and photodiode working point [11].

Based on this physical picture, prior research has shown that OED enables highly sensitive wavelength detection. Ge PN devices exhibit a large dispersion in the C- and L-bands, enabling precise wavelength monitoring and FBG interrogation using a single detector [12]. Electronic enhancement techniques later increased the sensitivity to the femtometer scale, and additional studies utilized OED for material sensing [13]. More recent work has revealed a strong OED in silicon photodiodes and solar cells, enabling low-cost spectral sensing in the visible and NIR [2]. Tunable OED in Ge PIN structures further enabled electrical control over the sign and magnitude of dispersion [9]. Recently, OED-based strain sensing has also been demonstrated in a Si photovoltaic cell [11], and a multi-physics simulation of OED in Si photodiodes has also been reported [10]. Collectively, these studies establish OED as a compact, versatile, and cost-effective mechanism for high-resolution optical sensing without the need for traditional dispersive elements or spectrometers.

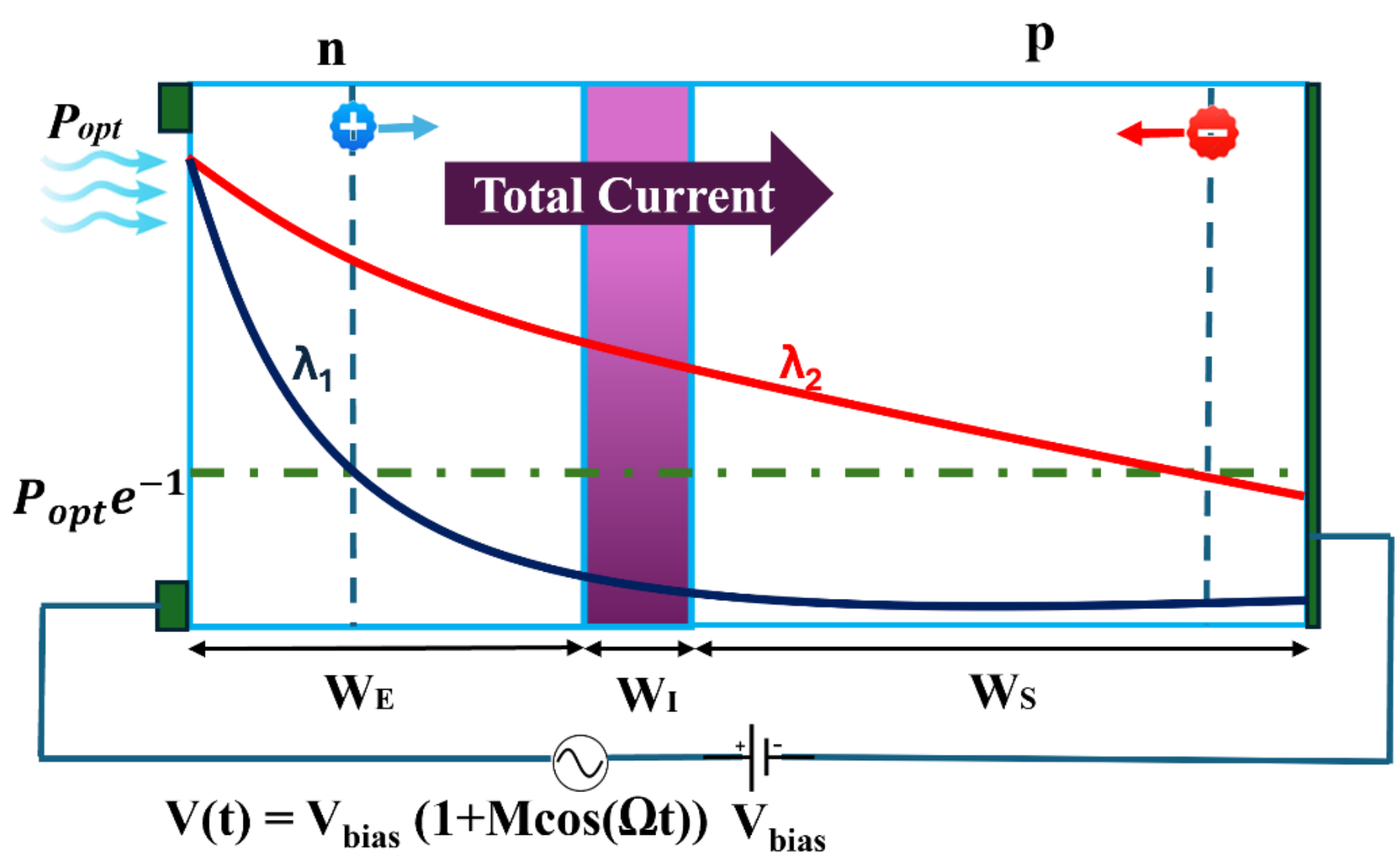

***Fig. 1:*** *Schematic of bm-OED in a P–N junction photodiode with CW illumination on the n-side.* $W_I$ *is the width of the depletion region and* $\lambda_2 > \lambda_1$.

In previous work, the OED in Ge and Si photodiodes was activated by modulating the incident light, which we refer to as light-modulated OED (lm-OED). Optical modulation also modulates the photocarrier generation rate, so the measured phase response is governed mainly by the generation and subsequent transport of the modulated carriers. The primary drawback of this method lies in the requirement of modulating the incoming radiation to the detector. This constraint restricts its applicability to specific scenarios where the light source is actively controlled. The technique presented here shifts the modulation from the optical radiation to the electronic modulation of the detector, which we call bias-modulated OED (bm-OED). This eliminates the primary constraint inherent in lm-OED, thereby enabling widespread implementation across any optical radiation and opening many applications in spectral sensing.

This work combines theory and experiment to investigate bm-OED in Ge PN photodiodes across the C- and L-bands (1525–1610 nm), examining their dependence on the wavelength, bias level, and optical power. Building on a theoretical model derived from the continuity equation, we calculate the internal current density, evaluate the harmonic components of the modulated response, and link them to changes in drift and space-charge screening. By varying the optical power and bias at a fixed RF power and frequency, the measured OED response is quantitatively consistent with these theoretical predictions [14]. We demonstrate how the screening effects and carrier transport dynamics jointly govern the modulation response. In the spectral range of 1525–1610 nm with a 3 MHz RF frequency, the OED sensitivity peaks at approximately 4 deg/nm, which is significantly higher than that of lm-OED in Ge. These results expand the physical understanding of bm-OED and outline a pathway toward actively tunable dispersive photodiodes for RF-photonics, wavelength-sensing, and optoelectronic signal processing. Potential applications include chip-scale computational spectrometers, photonic integrated circuits, portable, wearable, and IoT spectroscopy platforms, where machine learning decodes the OED signature to reconstruct optical spectra from a single low-cost detector [5].

## 2. Theory

Starting from the van Roosbroeck continuity equations [15,16], we derive the RF phase dependence on wavelength under bias modulation. Unlike light modulation, which primarily perturbs carrier generation under a fixed electrostatic potential, bias modulation directly drives the electric field in the depletion region of the photodiode. This dynamic field modulation alters carrier transport pathways and transit-time delays, with the wavelength-dependent absorption profile translating these changes into a bias-controlled phase response.

Fig. 1 displays an n-on-p bias-modulated photodiode with CW illumination on the n-side, giving a time-independent charge-carrier generation rate $g(x) = \alpha I_0(x)e^{-\alpha x}$ for an input optical intensity $I_0$. The transport calculation is done within the depletion region $W_E \leq x \leq W_E + W_I$, where drift current dominates under the influence of the built-in and applied fields.

The applied voltage $V_{\mathrm{app}}(t) = V_{bias}\,(1 + Mcos\,(\Omega t))$, where $M$ is the modulation parameter and $V_{bias}$ is the reverse bias voltage. In practice, at sufficiently high optical powers, photogenerated holes drift more slowly than electrons and accumulate in the depletion region, creating positive space charge that screens the electric field and slows carrier collection. Consequently, the net internal field is reduced below the value that the external bias alone would establish [17–20]. Accounting for this opposing screening contribution, the total voltage across the illuminated junction is expressed as

$$V_{tot}(t) = V_{bi,eff} + V_{app}(t) - V_{scr} = 2V_{bi,eff} - V_{bi,dark} + V_{app}(t), \quad (1)$$

where $V_{bi,eff}$ is the effective built-in voltage and $V_{scr} = V_{bi,dark} - V_{bi,eff}$ is the voltage reduction due to photocarrier-induced field screening within the depletion region. For the sake of simplicity, a uniform field is assumed throughout the depletion region, Correspondingly, the instantaneous electric field in the depletion region of width $W_I$ is $E(t) = \frac{V_{tot}(t)}{W_I}$. Neglecting recombination, the continuity equations for the electron (n) and hole (p) carrier densities are:

$$\frac{\partial n(x,t)}{\partial t} = g(x) + v_n(t)\,\frac{\partial n(x,t)}{\partial x}, \quad (2a)$$

$$\frac{\partial p(x,t)}{\partial t} = g(x) - v_p(t)\,\frac{\partial p(x,t)}{\partial x} \qquad (2b)$$

where the current velocities are

$$v_n(t) = \mu_n E(t) = \mu_n E_0(1 + M\cos(\Omega t)) = v_{n0}(1 + M\cos(\Omega t)) \qquad (3a)$$

$$v_p(t) = \mu_p E(t) = \mu_p E_0(1 + M\cos(\Omega t)) = v_{p0}(1 + M\cos(\Omega t)) \qquad (3b)$$

and the current densities are

$$J_n = qn\mu_n E, \qquad J_p = qp\mu_p E \qquad (4)$$

To derive the solutions for the currents, we assume separation of variables for the spatial and temporal description of the carrier densities, i.e. $n(x,t) = n_x(x) \cdot n_T(t)$ and $p(x,t) = p_x(x) \cdot p_T(t)$. Physically, the spatial factor reflects the intrinsic carrier distribution, fixed by doping profile and optical absorption, whereas the temporal factor represents the externally imposed modulation of this distribution. This assumption is certainly valid in the weak-modulation ($M \ll 1$) quasi-static regime.

Under these conditions, the solution for the electron density is

$$n(W_E,t) = \alpha I_0\, e^{\alpha W_E}\Bigg[\frac{1}{\alpha v_{n0}} - \frac{(\alpha v_{n0})^2(M/\Omega)^2}{(\alpha v_{n0})^2+\Omega^2}\cdot\frac{\alpha v_{n0}}{2} + \left\{\frac{(\alpha v_{n0})^2(M/\Omega)}{(\alpha v_{n0})^2+\Omega^2} - \frac{M}{\Omega}\right\}\sin(\Omega t) - \frac{(\alpha v_{n0})\left(\frac{M}{\Omega}\right)\Omega}{(\alpha v_{n0})^2+\Omega^2}\cos(\Omega t) + \frac{(\alpha v_{n0})^2(M/\Omega)^2}{(\alpha v_{n0})^2+\Omega^2}\cdot\frac{\Omega}{2}\sin(2\Omega t) + \frac{(\alpha v_{n0})^2(M/\Omega)^2}{(\alpha v_{n0})^2+\Omega^2}\cdot\frac{\alpha v_{n0}}{2}\cos(2\Omega t)\Bigg] \qquad (5)$$

and consists of dc, fundamental and second-harmonic terms.

This leads to the following expression for the electron current density, which contains an additional third-order harmonic term:

$$J_n(t) = q\, v_{n0}\,(1 + M\cos(\Omega t))n(W_E,t) = J_n^{dc} + J_n^{\Omega} + J_n^{2\Omega} + J_n^{3\Omega} \qquad (6)$$

where $J_n^{dc} = q\alpha v_{n0} I_0 e^{\alpha W_E}\left[\frac{1}{\alpha v_{n0}} - \frac{(\alpha v_{n0})^2(M/\Omega)^2}{(\alpha v_{n0})^2+\Omega^2}\cdot\frac{\alpha v_{n0}}{2} - \frac{M}{2}\frac{(\alpha v_{n0})(M/\Omega)\Omega}{(\alpha v_{n0})^2+\Omega^2}\right]$

$$J_n^{\Omega} = q\alpha v_{n0} I_0 e^{\alpha W_E}\Bigg[\left\{\frac{(\alpha v_{n0})^2(M/\Omega)}{(\alpha v_{n0})^2+\Omega^2} - \frac{M}{\Omega} + M\frac{(\alpha v_{n0})^2(M/\Omega)^2}{(\alpha v_{n0})^2+\Omega^2}\frac{\Omega}{4}\right\} Sin\Omega t + \left\{\frac{M\alpha v_{n0}}{(\alpha v_{n0})^2+\Omega^2} - M\frac{(\alpha v_{n0})^2(M/\Omega)^2}{(\alpha v_{n0})^2+\Omega^2}\frac{\alpha v_{n0}}{4} + \frac{M}{\alpha v_{n0}}\right\} Cos\Omega t\Bigg]$$

$$J_n^{2\Omega} = q\alpha v_{n0} I_0 e^{\alpha W_E}\Bigg[\left\{M\frac{(\alpha v_{n0})^2(M/\Omega)}{(\alpha v_{n0})^2+\Omega^2} - \frac{M^2}{2\Omega}\right\} Sin2\Omega t + \left\{\frac{(\alpha v_{n0})^2(M/\Omega)^2}{(\alpha v_{n0})^2+\Omega^2}\frac{\alpha v_{n0}}{2} - \frac{M}{2}\frac{(\alpha v_{n0})(M/\Omega)\Omega}{(\alpha v_{n0})^2+\Omega^2}\right\} Cos2\Omega t\Bigg]$$

$$J_n^{3\Omega} = q\alpha v_{n0} I_0 e^{\alpha W_E} M\left[\frac{(\alpha v_{n0})^2(M/\Omega)^2}{(\alpha v_{n0})^2+\Omega^2}\frac{\alpha v_{n0}}{4} Cos3\Omega t + \frac{(\alpha v_{n0})^2(M/\Omega)^2}{(\alpha v_{n0})^2+\Omega^2}\frac{\Omega}{4} Sin3\Omega t\right].$$

In the same fashion, we solve for the hole density at $x = W_E + W_I$ leading to

$$J_p(t) = q\, v_{p0}(1 + M\cos\Omega t)p(W_E + W_I, t) = J_p^{dc} + J_p^{\Omega} + J_p^{2\Omega} + J_p^{3\Omega} \qquad (7)$$

where $J_p^{dc} = q\alpha v_{p0} I_0 e^{-\alpha(W_E+W_I)} \left[\frac{(\alpha v_{p0})^2 (M/\Omega)^2}{(\alpha v_{p0})^2+\Omega^2} \cdot \frac{\alpha v_{p0}}{2} - \frac{1}{\alpha v_{p0}} + \frac{M}{2} \frac{(\alpha v_{p0})(M/\Omega)\Omega}{(\alpha v_{p0})^2+\Omega^2}\right]$

$$J_p^{\Omega} = q\alpha v_{p0} I_0 e^{-\alpha(W_E+W_I)} \left[\left\{\frac{(\alpha v_{p0})^2 (M/\Omega)}{(\alpha v_{p0})^2+\Omega^2} - \frac{M}{\Omega} + M \frac{(\alpha v_{p0})^2 (M/\Omega)^2}{(\alpha v_{p0})^2+\Omega^2} \frac{\Omega}{4}\right\} Sin\Omega t + \left\{\frac{(\alpha v_{p0})(M/\Omega)\Omega}{(\alpha v_{p0})^2+\Omega^2} - \frac{M}{\alpha v_{p0}} + M \frac{(\alpha v_{p0})^2 (M/\Omega)^2}{(\alpha v_{p0})^2+\Omega^2} \frac{\alpha v_{p0}}{4}\right\} Cos\Omega t\right]$$

$$J_p^{2\Omega} = q\alpha v_{p0} I_0 e^{-\alpha(W_E+W_I)} \left[\left\{M \frac{(\alpha v_{p0})^2 (M/\Omega)}{(\alpha v_{p0})^2+\Omega^2} - \frac{M^2}{2\Omega}\right\} Sin2\Omega t + \left\{\frac{M}{2} \frac{(\alpha v_{p0})(M/\Omega)\Omega}{(\alpha v_{p0})^2+\Omega^2} - \frac{(\alpha v_{p0})^2 (M/\Omega)^2}{(\alpha v_{p0})^2+\Omega^2} \frac{\alpha v_{p0}}{2}\right\} Cos2\Omega t\right]$$

$$J_p^{3\Omega} = q\alpha v_{p0} I_0 e^{-\alpha(W_E+W_I)} M \left[\frac{(\alpha v_{p0})^2 (M/\Omega)^2}{(\alpha v_{p0})^2+\Omega^2} \frac{\Omega}{4} Sin3\Omega t - \frac{(\alpha v_{p0})^2 (M/\Omega)^2}{(\alpha v_{p0})^2+\Omega^2} \frac{\alpha v_{p0}}{4} Cos3\Omega t\right]$$

The total drift current density is

$$J_{\text{drift}}(t) = J_p(t) + J_n(t), \qquad (8)$$

The source of the nonlinearity in the carrier density dynamics is the coupling of the sinusoidally modulated carrier velocity with the exponential depth-dependent photocarrier-generation profile (eq. 2). The source of the additional nonlinearity in the current density is due to the multiplication of the modulated carrier velocity by the nonlinear time-dependent carrier density (eq. 4). Obviously, stronger modulation will introduce higher harmonic terms. Because the absorption depth, and therefore modulation, varies across the optical spectrum, these harmonic amplitudes and phases become wavelength-dependent, giving rise to the nonlinear response characteristic of bm-OED. This interpretation is consistent with established photodiode nonlinearity studies, which attribute harmonic distortion to field-dependent carrier transport and space-charge effects [21,22]. The diffusion current is excluded because it depends on the carrier concentration gradient set by optical generation, which is not influenced by the applied bias. In contrast, drift and displacement currents directly follow from the field and its time derivative.

The total current density is therefore

$$J_{\text{total}}(t) = J_{drift}(t) + J_d(t), \qquad (9)$$

where $J_d(t)$ accounts for a displacement current density

$$J_d(t) = -\varepsilon_r \varepsilon_0 \, E_0 \, M \, \Omega \sin(\Omega t). \qquad (10)$$

where $E_0 = \frac{V_{bi,eff} + V_{bias} - V_{scr}}{W_I}$ is the dc electric field.

From Eq. 9, the OED sensitivity $S_{OED}$ is derived and plotted together with experimental results, as discussed in the following section.

## 3. Experimental Setup

The schematic of the experimental system is shown in Fig. 2. CW light from a narrow-linewidth (25kHz) tunable laser (ID Photonics CoBrite DX) with tuning over the C- and L-bands, illuminated a Ge PN-type photodiode (GPD Optoelectronics GM3). A function generator (Agilent 33621A) supplied the reverse-bias to the photodiode, and the photocurrent-induced voltage was measured across a 50 Ω load resistor. The photodiode output current was measured using a vector network analyzer (Rohde and Schwarz ZNLE4) to obtain RF amplitude and phase information.

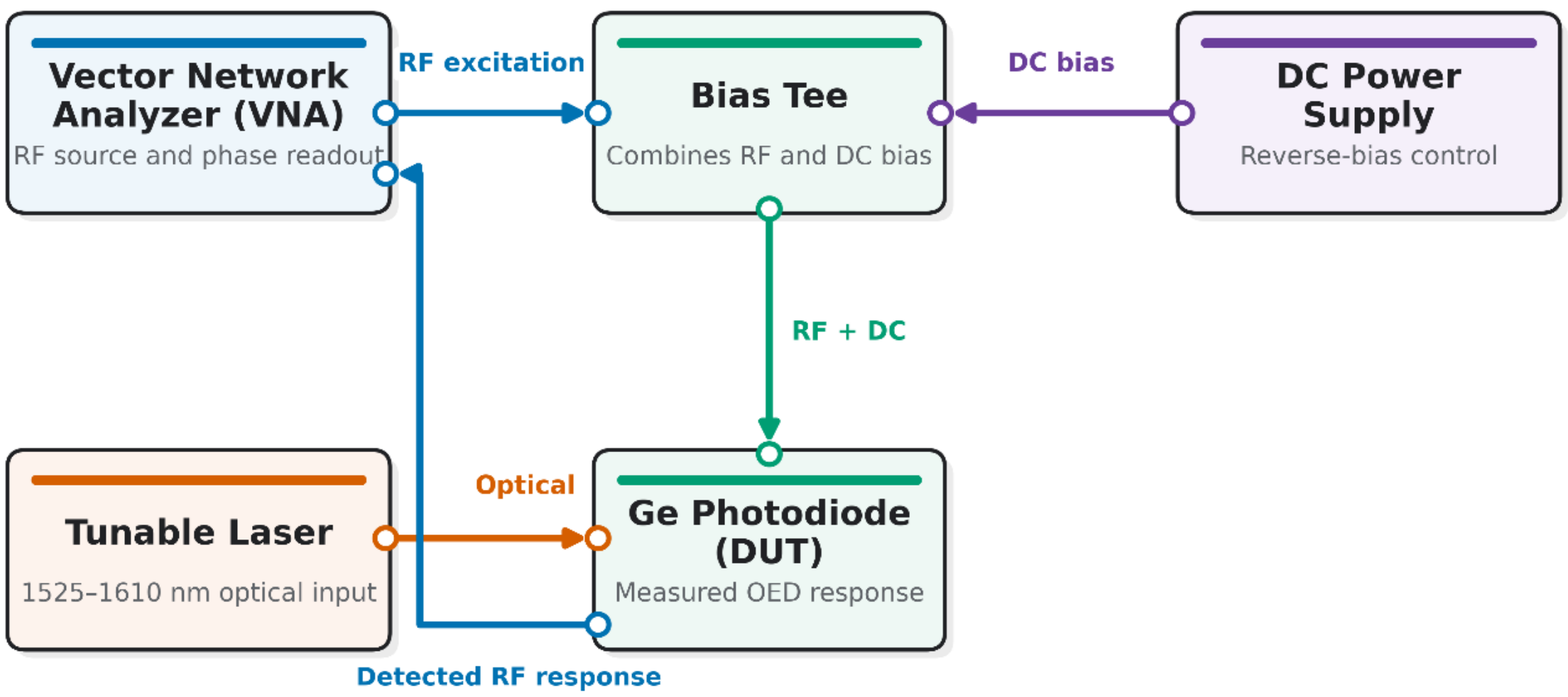

***Fig. 2:*** *Schematic of the experimental setup. A tunable laser illuminates the photodiode, while a vector network analyzer (VNA) supplies the RF modulation and measures the S-parameter response. The RF and DC signals are combined via a bias tee, with the DC bias supplied by a DC power supply, enabling simultaneous RF phase measurement and bias modulation of the device.*

The tunable laser was initialized at 1525 nm and swept to 1610 nm in 5 nm increments, while the bias voltage was scanned from 1 V to 10 V in steps of 1 V. All measurements were performed at 3 MHz, since the OED was strongest in this region. The experiment was performed in a room-temperature-controlled environment to eliminate temperature-dependent phase drifts in the optical and RF components.

## 4. Results and discussion

Fig. 3 displays plots of the predicted and measured OED sensitivity as a function of wavelength across bias voltages ranging from 1 to 10 V, for optical powers between 0.2 and 5.7 mW. The following values were used for the various parameters: active illumination diameter of 100 µm, $\varepsilon_r$ =16.2, $f$=3 MHz, and built-in dark voltage ($V_{bi,dark}$)=0.28 V, with nominal values for the widths $W_E$, $W_I$ and $W_S$ and the mobilities $\mu_e$ and $\mu_h$. The values of $V_{tot}$ were determined from eq. 1 by measuring the *C-V* response of the detector for varying optical power and extracting the optical power-dependent $V_{scr}$ from these graphs, as explained in the Supplementary Section.

Fig. 3 displays regions where the predictions and the data are in agreement, as well as other regions where there is some divergence. The spectral response is due to wavelength-dependent optical absorption together with bias-dependent competition between carrier drift, displacement current, and mobile-charge screening. Given the complex interaction between these effects, the simplifying assumptions of the model, and the lack of information regarding the actual values of the physical parameters of the photodiode, it is challenging to achieve good agreement across the spectrum.

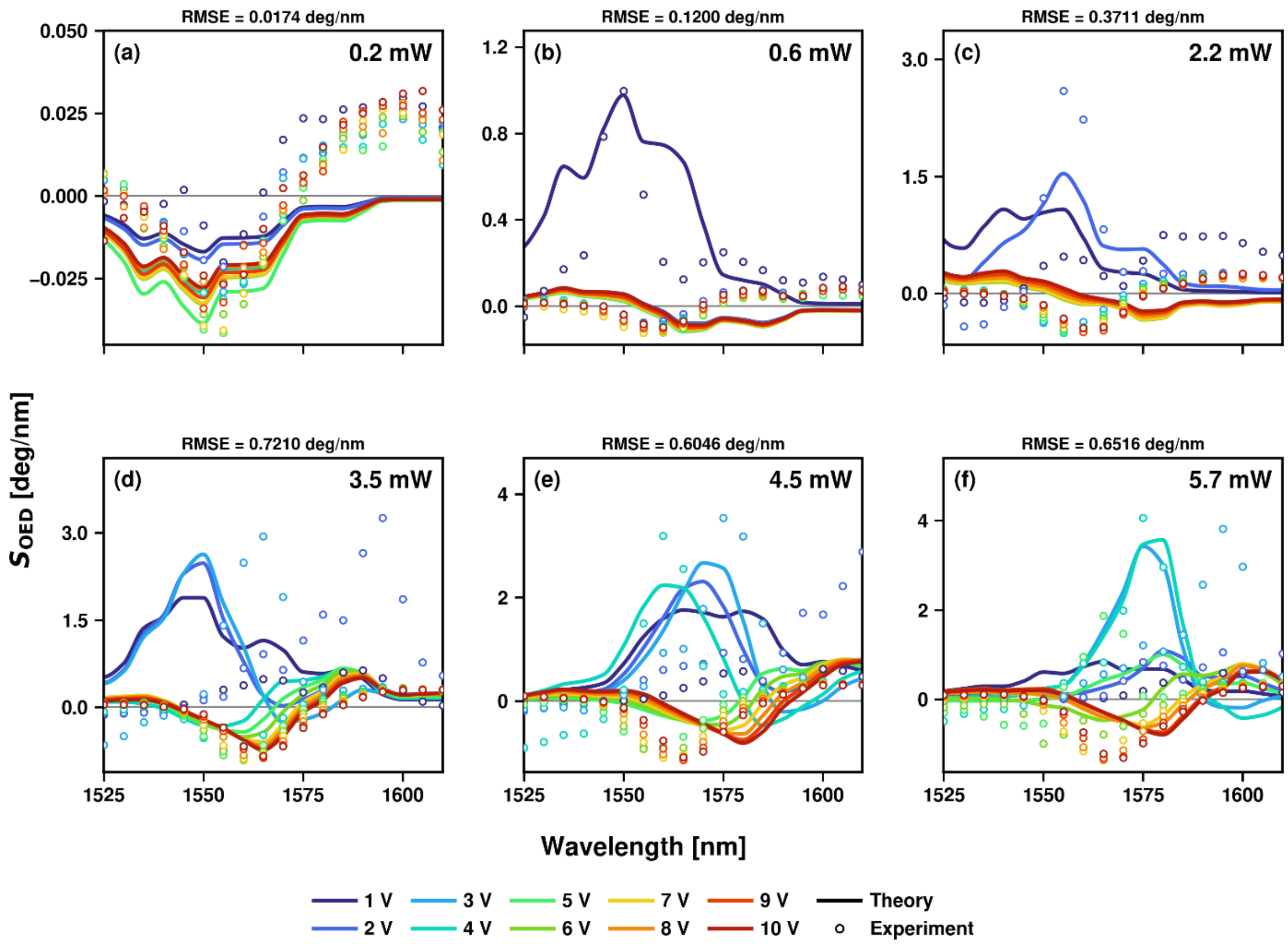

*Fig. 3: **OED sensitivity vs. wavelength for various optical powers.** Experimental OED sensitivities (open circles) and theoretical results (solid lines) are shown for reverse biases of 1–10 V at optical powers of (a) 0.2 mW, (b) 0.6 mW, (c) 2.2 mW, (d) 3.5 mW, (e) 4.5 mW and (f) 5.7 mW. The theory reproduces the bias-dependent evolution of the negative and positive spectral features with increasing optical power.*

Overall, we can identify several features of the experimental data that merits discussion:

**1. Fig. 3a: Low optical power regime**

At 0.2 mW optical power, the OED sensitivity is predominantly negative in the C-band and positive for the longer wavelength L-band. This is due to the relative contribution of the hole and electron current to the overall response. In the C-band, the light penetration depth is relatively shallow, so that the predominant carriers are the holes formed in the n-type entrance region. Electron carriers are also generated in the deeper tail region of the illumination, however due to the weaker light power in this region, their generation rate is weaker. In addition, the hole velocity is smaller than that of the electron velocity. Therefore, overall, the OED is predominantly characterized by the phase delay of the hole current. As the wavelength increases, the carrier generation region approaches the junction, shortening the hole path and reducing the phase delay. Therefore, the OED sensitivity, which we recall is defined as $\frac{d\theta}{d\lambda}$, is predominantly negative in this spectral region. In the L-band, the electron current begins to dominate, so that the situation is reversed – as the wavelength increases, the electron transit time increases, and the OED sensitivity is therefore positive.

Voltage dependence provides an additional modification to this spectral behavior. At 1 V, the junction is depleted, and both carrier velocities remain in the low-field regime. As the bias increases to 5–6 V, the stronger electric field shortens the carrier transit times. In the C-band, the changing depletion width increases the wavelength-dependent difference between the electron and hole delays, so the phase decreases more rapidly with wavelength, and the negative OED sensitivity increases in magnitude. In the L-band, the stronger field reduces the additional delay of carriers generated deeper in the device; therefore, the phase still increases with wavelength, but at a lower rate, giving a smaller positive sensitivity. At 10 V, the junction is nearly fully depleted, so the depletion width no longer changes significantly. The electron velocity also approaches its high-field limit earlier than the hole velocity. Consequently, the C-band phase becomes less dependent on wavelength and its negative sensitivity weakens. In the L-band, the positive electron-delay contribution hardly changes, whereas the opposing hole-delay contribution becomes smaller as the holes continue to accelerate. The electron contribution therefore becomes dominant again, causing the positive OED sensitivity to increase, producing the peak positive OED sensitivity of +0.032 deg/nm.

**2. Fig. 3b-f: Increasing optical power**

As the power increases, there is an increasing trend for positive OED sensitivity peaks to appear, first at lower bias and then for higher bias. This is due to the role of the increasing screening voltage. With increasing light power, the carrier generation rate increases. At low reverse bias, carrier extraction is slow, and the junction field drives electrons toward the n-side and holes toward the p-side, producing spatially separated screening charge. The resulting screening field opposes the junction field and reduces the effective field, which further slows extraction, leading to further persistence of the screening charge [23–25]. With higher optical power, this screening-related relaxation strengthens and appears over an increasing number of spectral regions at low bias.

The maximum positive OED sensitivity rises from 1 deg/nm at 0.6 mW to +4.13 deg/nm at 5.7 mW, while the largest negative sensitivity grows from −0.124 to −1.364 deg/nm. At higher reverse bias, the depletion region widens and carrier extraction speed increases, which reduces the screening charge and weakens its relaxation contribution. In this case, the underlying drift response dominates again: in the C-band, increasing wavelength shortens the transport path of the slower holes, giving negative OED, while in the L-band, deeper photogeneration increases the electron transport path, giving positive OED. As the optical power increases further, the larger photocarrier population requires progressively higher reverse bias to overcome screening [26], shifting the OED transition to higher voltages.

These effects combine to produce a maximum OED parameter in the L-band (Fig. 3f, where $S_{OED}$ = 4.13 deg/nm) of $D_{OED} = 5.46 \times 10^{10}$ ps/(nm·km) at 3 MHz modulation, assuming $W$ = 70 μm. This exceeds our previously reported dispersion for light-modulated Ge PN OED by several orders of magnitude [3,12,13], and is more than nine orders of magnitude above standard SMF-28 fiber (17 ps/(nm·km)) [27], compressing the dispersive effect of tens of thousands of kilometers of fiber into tens of micrometers.

Compared with conventional lm-OED, which probes wavelength-dependent carrier generation under a fixed electric field, bm-OED achieves a seven-fold enhancement in sensitivity by directly probing electric-field-controlled carrier transport and transit dynamics. The qualitative agreement between theory and experiment across a wide range of optical power demonstrates that the model captures the dominant wavelength, bias, and power dependent transport mechanisms, though some quantitative deviations remain for the reasons we discussed above. In addition, the present analysis is limited to a 1D treatment; a full 3D TCAD physics simulation is currently under development [10].

Experimentally, the 3-sigma phase noise of approximately 15 mdeg at 500 Hz- bandwidth supports wavelength monitoring with a resolution of 3.6 pm. More broadly, the electrically programmable magnitude and sign of the OED response establish the photodiode as a phase-domain computational element, providing a potential route towards photodiode-only computational spectroscopy [28,29], in-sensor spectral intelligence [29,30], computational hyperspectral/4D imaging [31,32], and photonic reservoir computing [33] in which detection, spectral weighting and temporal memory are integrated within a single junction.

## 5. Conclusions

This study presents a theoretical model and experimental demonstration of bias-modulated optoelectronic chromatic dispersion in a commercial Ge PN photodiode, where reverse-bias modulation under constant continuous-wave illumination produces a wavelength-dependent RF phase response. A peak sensitivity of +4.13 deg/nm was measured at 5.7 mW and 4 V, and an effective dispersion parameter of $5.46 \times 10^{10}$ ps/(nm·km) when normalized to the 70 μm device length. This value is many orders of magnitude greater than that of standard optical materials such as single-mode fiber, and several-fold larger than previously reported light-modulated OED values. The 15 mdeg phase-noise floor corresponds to a projected wavelength resolution of approximately 3.6 pm. The close agreement between theory and experiment attributes this behavior to the bias- and power-dependent competition among drift, displacement and delayed admittance responses, with screening modifying their relative contributions by reducing the internal electric field.

These results motivate three research directions. First, Ge PIN structures with controlled intrinsic-layer thickness and doping could provide systematic engineering of the trade-off among OED sensitivity, electrical bandwidth and unambiguous spectral range. Second, extending bm-OED to Si and InGaAs photodiodes and photonic integrated circuits could enable electrically programmable wavelength-to-phase conversion for microwave-photonic processing. Third, arrays of independently biased photodiodes could provide signed and reconfigurable spectral-phase kernels for filter-free computational spectroscopy and in-sensor photonic processing. More broadly, bm-OED transforms reverse bias from a static operating condition into an active control parameter for programming the magnitude, sign, and spectral position of the photodetector phase response.

## References

1. F. H. L. Koppens, T. Mueller, Ph. Avouris, A. C. Ferrari, M. S. Vitiello, and M. Polini, "Photodetectors based on graphene, other two-dimensional materials and hybrid systems," Nat. Nanotechnol. **9**, 780–793 (2014).
2. S. Mudgal, P. K. Dubey, Z. Glasser, and S. Sternklar, "Large optoelectronic chromatic dispersion in PN-type silicon photodiodes and photovoltaic cells," Opt. Lett. **49**, 2185 (2024).
3. Z. Glasser, P. K. Dubey, A. Dutta, E. Liokumovitch, R. Abramov, and S. Sternklar, "Optoelectronic chromatic dispersion and wavelength monitoring in a photodiode," Opt. Express **29**, 19839–19852 (2021).
4. E. E. Kassa, Z. Glasser, U. K. Saint, R. Yozevitch, and S. Sternklar, "Optoelectronic Chromatic Dispersion in a Single Photodiode for Machine-Learning-Based Computational Spectroscopy," (2026).
5. E. E. Kassa, Z. Glasser, U. K. Saint, R. Yozevitch, and S. Sternklar, "A Photodiode-Only Platform for Machine-Learning-Based Computational Spectroscopy Based on Optoelectronic Chromatic Dispersion," in *CLEO 2026* (Optica Publishing Group, 2026), p. AW2B.1.
6. Z. Wang, S. Wang, B. Gao, C. Li, and W. Yu, "Real-time spectral reconstruction method with noise robustness based on a 2D-CNN network for snapshot spectral imaging sensors," Opt. Express **33**, 54358 (2025).
7. S. Wang, H. Yin, Y. Li, Z. Du, Y. Zhang, and S. X. Zhang, "Spectrum Reconstruction Model Based on Multispectral Electrochromic Devices," Adv. Sci. **11**, 2400979 (2024).
8. J. Zhang, R. Su, Q. Fu, W. Ren, F. Heide, and Y. Nie, "A survey on computational spectral reconstruction methods from RGB to hyperspectral imaging," Sci. Rep. **12**, 11905 (2022).
9. A. Dutta, E. Liokumovitch, Z. Glaser, and S. Sternklar, "Large and tunable optoelectronic chromatic dispersion in PIN-type photodiodes," Opt. Lett. **49**, 2057 (2024).
10. M. Itzhaki, P. Medved, M. Engel, Z. Glasser, S. Sternklar, and A. Albo, "Predictive multi-physics modeling of optoelectronic chromatic dispersion in silicon photodiodes," Opt. Express **34**, 31677 (2026).
11. A. Dutta, S. Mudgal, Z. Glasser, E. Liokumovitch, O. Newman, M. Shuldiner, and S. Sternklar, "Universal phase-based self-powered strain sensing in photovoltaic cells enabled by optoelectronic chromatic dispersion," Sens. Actuators Phys. **410**, 118315 (2026).
12. E. Liokumovitch, Z. Glasser, L. Singh, R. Abramov, and S. Sternklar, "Optoelectronic chromatic dispersion in germanium PN photodiodes: wavelength monitoring and FBG interrogation," Opt. Lett. **46**, 4061 (2021).
13. E. Liokumovitch, Z. Glasser, and S. Sternklar, "Femtometer-resolved wavelength monitor based on photodiode optoelectronic chromatic dispersion with RF phase-shift amplification," Opt. Lett. **47**, 2622 (2022).
14. U. K. Saint, Z. Glasser, E. E. Kassa, and S. Sternklar, "Bias-Modulated Optoelectronic Chromatic Dispersion in Germanium PN Photodiodes," in *CLEO 2026* (Optica Publishing Group, 2026), p. STH1G.2.
15. W. Van Roosbroeck, "Theory of the Flow of Electrons and Holes in Germanium and Other Semiconductors," Bell Syst. Tech. J. **29**, 560–607 (1950).
16. W. Van Roosbroeck, "The Transport of Added Current Carriers in a Homogeneous Semiconductor," Phys. Rev. **91**, 282–289 (1953).
17. X.-G. Zhang and S. T. Pantelides, "Theory of Space Charge Limited Currents," Phys. Rev. Lett. **108**, 266602 (2012).
18. M. Sajedi Alvar, P. W. M. Blom, and G.-J. A. H. Wetzelaer, "Space-charge-limited electron and hole currents in hybrid organic-inorganic perovskites," Nat. Commun. **11**, 4023 (2020).
19. D.-P. Han, M. Iwaya, T. Takeuchi, S. Kamiyama, and I. Akasaki, "Space-charge effect on photogenerated-current and -voltage in III-nitride optoelectronic semiconductors," Photonics Res. **9**, 1820 (2021).
20. S. N. Mohammad and S. T. H. Abidi, "Theory of saturation photocurrent and photovoltage in *p* - *n* junction solar cells," J. Appl. Phys. **61**, 4909–4919 (1987).
21. R. R. Hayes and D. L. Persechini, "Nonlinearity of p-i-n photodetectors," IEEE Photonics Technol. Lett. **5**, 70–72 (1993).
22. K. J. Williams, R. D. Esman, and M. Dagenais, "Nonlinearities in p-i-n microwave photodetectors," J. Light. Technol. **14**, 84–96 (1996).
23. C. E. Campanella, A. Cuccovillo, C. Campanella, A. Yurt, and V. M. N. Passaro, "Fibre Bragg Grating Based Strain Sensors: Review of Technology and Applications," Sensors **18**, 3115 (2018).
24. X. Cui, F. Nigmatulin, L. Wang, I. Reduto, A. C. Liapis, M. Du, M. G. Uddin, S. Abde Mayeen, F. Ahmed, Y. Zhang, H. H. Yoon, H. Lipsanen, S. Honkanen, T. Aalto, Z. Yang, T. Hasan, W. Cai, and Z. Sun, "Miniaturized spectral sensing with a tunable optoelectronic interface," Sci. Adv. **11**, eado6886 (2025).
25. D. Kwak, D. K. Polyushkin, and T. Mueller, "In-sensor computing using a MoS2 photodetector with programmable spectral responsivity," Nat. Commun. **14**, 4264 (2023).
26. A. Beling, X. Xie, and J. C. Campbell, "High-power, high-linearity photodiodes," Optica **3**, 328 (2016).
27. C. E. Campanella, A. Cuccovillo, C. Campanella, A. Yurt, and V. M. N. Passaro, "Fibre Bragg Grating Based Strain Sensors: Review of Technology and Applications," Sensors **18**, 3115 (2018).
28. X. Cui, F. Nigmatulin, L. Wang, I. Reduto, A. C. Liapis, M. Du, M. G. Uddin, S. Abde Mayeen, F. Ahmed, Y. Zhang, H. H. Yoon, H. Lipsanen, S. Honkanen, T. Aalto, Z. Yang, T. Hasan, W. Cai, and Z. Sun, "Miniaturized spectral sensing with a tunable optoelectronic interface," Sci. Adv. **11**, eado6886 (2025).
29. D. Kwak, D. K. Polyushkin, and T. Mueller, "In-sensor computing using a MoS2 photodetector with programmable spectral responsivity," Nat. Commun. **14**, 4264 (2023).
30. D. Zhang, Y. Li, J. Geng, H. M. Kim, M. Ma, S. Wang, I. Kim, T. J. Wijaya, N. Higashitarumizu, I. K. M. R. Rahman, D. Urmossy, J. Bullock, A. Ozcan, and A. Javey, "Spectral kernel machines with electrically tunable photodetectors," Science **390**, eady6571 (2025).
31. F. F. Settembrini, A. C. Gungor, A. Forrer, S. A. Fortune, A. Dell'Aquila, P. Padmanabhan, I. E. Opris, M. Sotto, N. Dordevic, Y. Perelman, T. Christen, M. Wang, and R. Nicolaescu, "A large-scale coherent 4D imaging sensor," Nature **651**, 364–370 (2026).
32. L. Bian, Z. Wang, Y. Zhang, L. Li, Y. Zhang, C. Yang, W. Fang, J. Zhao, C. Zhu, Q. Meng, X. Peng, and J. Zhang, "A broadband hyperspectral image sensor with high spatio-temporal resolution," Nature **635**, 73–81 (2024).

33. A. Aadhi, L. Di Lauro, B. Fischer, P. Dmitriev, I. Alamgir, C. Mazoukh, N. Perron, E. A. Viktorov, A. V. Kovalev, A. Eshaghi, S. Vakili, M. Chemnitz, P. Roztocki, B. E. Little, S. T. Chu, D. J. Moss, and R. Morandotti, "Scalable photonic reservoir computing for parallel machine learning tasks," Nat. Commun. **17**, 1225 (2025).